\documentclass[preprint]{vgtc}             %

\ifpdf%
  \pdfoutput=1\relax                   %
  \usepackage{graphicx}                %
  \DeclareGraphicsExtensions{.pdf,.png,.jpg,.jpeg} %
\else%
  \usepackage{graphicx}                %
  \DeclareGraphicsExtensions{.eps}     %
\fi%

\graphicspath{{figures/}{pictures/}{images/}{./}} %

\usepackage{microtype}                 %
\PassOptionsToPackage{warn}{textcomp}  %
\usepackage{textcomp}                  %
\usepackage{mathptmx}                  %
\usepackage{times}                     %
\usepackage{cite}                      %
\usepackage{tabu}                      %
\usepackage{booktabs}                  %
\onlineid{0}

\vgtccategory{Research}

\vgtcinsertpkg

\renewcommand\footnotemark{}

\title{\tool{}: Exploring and Curating Sparse Decision Trees with Interactive Visualization}

\newcommand{\authorgap}{\hspace{10pt}}

\author{
  Zijie J. Wang\textsuperscript{\textrm 1} %
  \thanks{\textsuperscript{\textrm 1}Georgia Institute of Technology. \{\href{mailto:jayw@gatech.edu}{jayw}$\mid$\href{mailto:polo@gatech.edu}{polo}\}@gatech.edu} \authorgap
  Chudi Zhong\textsuperscript{\textrm 2} %
  \thanks{\textsuperscript{\textrm 2}Duke University. \{\href{mailto:chudi.zhong@duke.edu}{chudi.zhong}$\mid$\href{mailto:rui.xin926@duke.edu}{rui.xin926}$\mid$\href{mailto:zhi.chen1@duke.edu}{zhi.chen1}\}@duke.edu, \newline \textcolor{white}{.} \hspace{11pt} \href{mailto:cynthia@cs.duke.edu}{cynthia@cs.duke.edu}} \authorgap
  Rui Xin\textsuperscript{\textrm 2} \authorgap
  Takuya Takagi\textsuperscript{\textrm 3} %
  \thanks{\textsuperscript{\textrm 3}Fujitsu Laboratories. \href{mailto:takagi.takuya@fujitsu.com}{takagi.takuya@fujitsu.com}} \authorgap
  Zhi Chen\textsuperscript{\textrm 2} \authorgap \\
  Duen Horng Chau\textsuperscript{\textrm 1} \authorgap
  Cynthia Rudin\textsuperscript{\textrm 2} \authorgap
  Margo Seltzer\textsuperscript{\textrm 4} %
  \thanks{\textsuperscript{\textrm 4}University of British Columbia. \href{mailto:mseltzer@cs.ubc.ca}{mseltzer@cs.ubc.ca}} \authorgap
}

\teaser{
  \centering
  \vspace{-10pt}
  \setlength{\belowcaptionskip}{4pt}
  \setlength{\abovecaptionskip}{5pt}
  \includegraphics[width=1\textwidth]{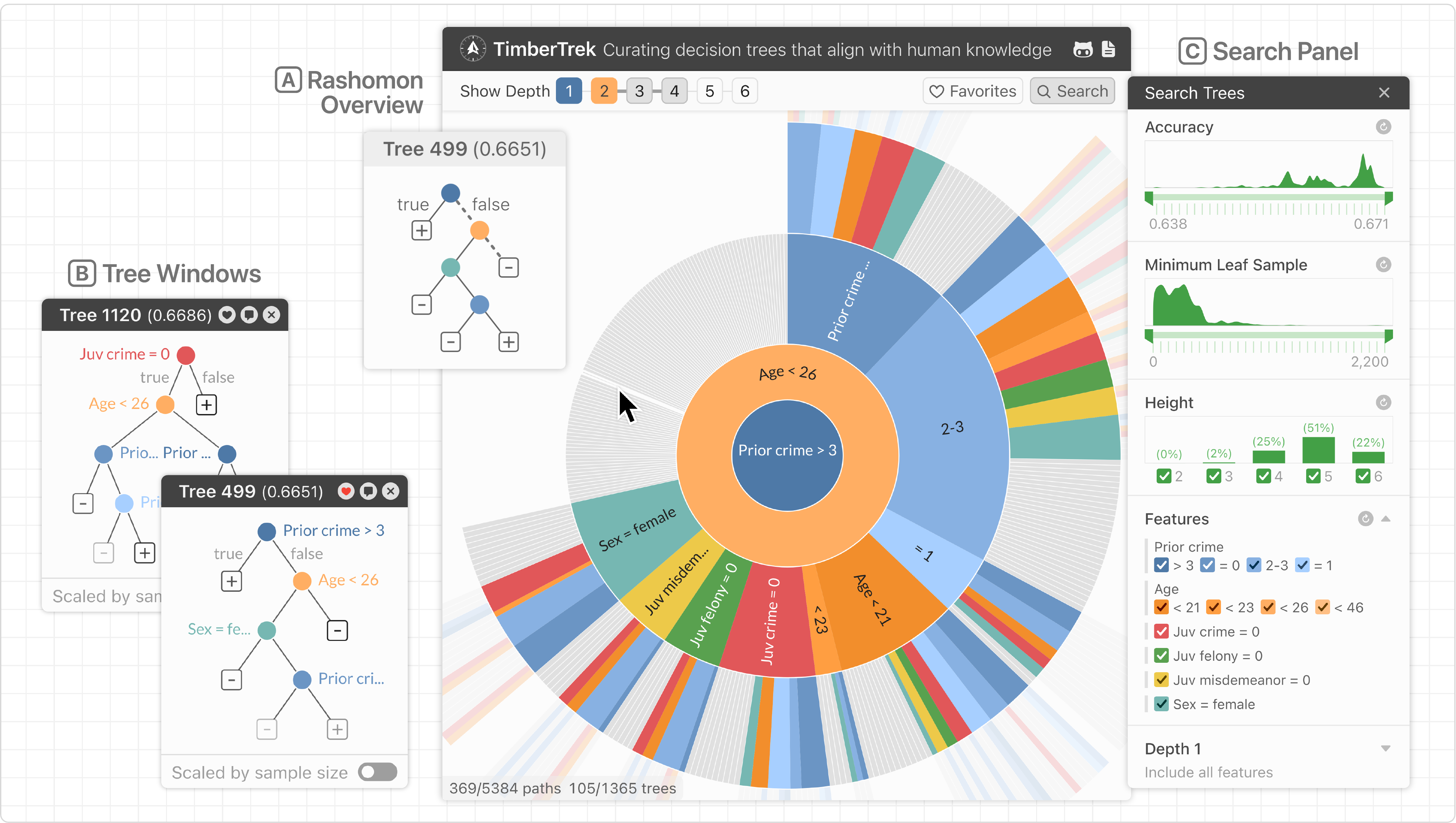}
  \caption{
    \tool{} empowers domain experts and data scientists to easily explore thousands of well-performing decision trees so they can find and collect those trees that best reflect their knowledge and values.
    Consider the task of predicting whether a criminal is likely to commit a crime in the next two years.
    \textbf{(A)} The \summaryview{} visually summarizes all well-performing decision trees by organizing them based on their decision paths, enabling users to seamlessly transition across different model subsets and explore trees with similar prediction patterns.
    \textbf{(B)} Clicking a tree opens a repositionable \treeview{} showing details of a decision tree: multiple windows allow users to compare several model candidates' prediction patterns.
    \textbf{(C)} The \searchview{} provides filtering tools, enabling users to quickly identify decision trees with desired properties, such as accuracy, robustness, simplicity, and used features.\looseness=-1
  }
  \label{fig:teaser}
  \setlength{\belowcaptionskip}{0pt}
  \setlength{\abovecaptionskip}{10pt}
}

\abstract{
Given thousands of equally accurate machine learning (ML) models, how can users choose among them?
A recent ML technique enables domain experts and data scientists to generate a complete \textit{Rashomon set} for sparse decision trees---a huge set of almost-optimal interpretable ML models.
To help ML practitioners identify models with desirable properties from this Rashomon set, we develop \tool{}, the first interactive visualization system that summarizes thousands of sparse decision trees at scale.
Two usage scenarios highlight how \tool{} can empower users to easily explore, compare, and curate models that align with their domain knowledge and values.
Our open-source tool runs directly in users' computational notebooks and web browsers, lowering the barrier to creating more responsible ML models.
\tool{} is available at the following public demo link: \link{https://poloclub.github.io/timbertrek}.
} 

\CCScatlist{
  \CCScatTwelve{Human-centered computing}{Visual Analytics}{}{}
}

\usepackage{enumitem}
\usepackage{tabu}
\usepackage{rotating}
\usepackage{tikz}
\usepackage{capt-of,etoolbox}
\usepackage{microtype}
\usepackage{bm}
\usepackage{soul}
\usepackage{wrapfig}
\usepackage{graphbox}
\usepackage{tcolorbox}
\usepackage{amssymb}
\usepackage{amsmath}
\usepackage{cleveref}
\usepackage{mathtools}
\usepackage{balance}
\usepackage[varqu]{zi4}
\usepackage[bb=boondox]{mathalfa}
\usepackage[utf8x]{inputenc}
\usepackage{units}
\usepackage{setspace}
\usepackage{gensymb}
\usepackage{helvet}
\usepackage{printlen} %
\usepackage[numbers]{natbib}

\usepackage{eqnarray}
\usepackage{bbm}

\definecolor{redOV}{RGB}{255, 235, 238}
\definecolor{redI}{RGB}{255, 205, 210}
\definecolor{redII}{RGB}{239, 154, 154}
\definecolor{redIII}{RGB}{229, 115, 115}
\definecolor{redIV}{RGB}{239, 83, 80}
\definecolor{redV}{RGB}{244, 67, 54}
\definecolor{redVI}{RGB}{229, 57, 53}
\definecolor{redVII}{RGB}{211, 47, 47}
\definecolor{redVIII}{RGB}{198, 40, 40}
\definecolor{redIX}{RGB}{183, 28, 28}
\definecolor{redAI}{RGB}{255, 138, 128}
\definecolor{redAII}{RGB}{255, 82, 82}
\definecolor{redAIV}{RGB}{255, 23, 68}
\definecolor{redAVII}{RGB}{213, 0, 0}

\definecolor{pinkOV}{RGB}{252, 228, 236}
\definecolor{pinkI}{RGB}{248, 187, 208}
\definecolor{pinkII}{RGB}{244, 143, 177}
\definecolor{pinkIII}{RGB}{240, 98, 146}
\definecolor{pinkIV}{RGB}{236, 64, 122}
\definecolor{pinkV}{RGB}{233, 30, 99}
\definecolor{pinkVI}{RGB}{216, 27, 96}
\definecolor{pinkVII}{RGB}{194, 24, 91}
\definecolor{pinkVIII}{RGB}{173, 20, 87}
\definecolor{pinkIX}{RGB}{136, 14, 79}
\definecolor{pinkAI}{RGB}{255, 128, 171}
\definecolor{pinkAII}{RGB}{255, 64, 129}
\definecolor{pinkAIV}{RGB}{245, 0, 87}
\definecolor{pinkAVII}{RGB}{197, 17, 98}

\definecolor{purpleOV}{RGB}{243, 229, 245}
\definecolor{purpleI}{RGB}{225, 190, 231}
\definecolor{purpleII}{RGB}{206, 147, 216}
\definecolor{purpleIII}{RGB}{186, 104, 200}
\definecolor{purpleIV}{RGB}{171, 71, 188}
\definecolor{purpleV}{RGB}{156, 39, 176}
\definecolor{purpleVI}{RGB}{142, 36, 170}
\definecolor{purpleVII}{RGB}{123, 31, 162}
\definecolor{purpleVIII}{RGB}{106, 27, 154}
\definecolor{purpleIX}{RGB}{74, 20, 140}
\definecolor{purpleAI}{RGB}{234, 128, 252}
\definecolor{purpleAII}{RGB}{224, 64, 251}
\definecolor{purpleAIV}{RGB}{213, 0, 249}
\definecolor{purpleAVII}{RGB}{170, 0, 255}

\definecolor{deeppurpleOV}{RGB}{237, 231, 246}
\definecolor{deeppurpleI}{RGB}{209, 196, 233}
\definecolor{deeppurpleII}{RGB}{179, 157, 219}
\definecolor{deeppurpleIII}{RGB}{149, 117, 205}
\definecolor{deeppurpleIV}{RGB}{126, 87, 194}
\definecolor{deeppurpleV}{RGB}{103, 58, 183}
\definecolor{deeppurpleVI}{RGB}{94, 53, 177}
\definecolor{deeppurpleVII}{RGB}{81, 45, 168}
\definecolor{deeppurpleVIII}{RGB}{69, 39, 160}
\definecolor{deeppurpleIX}{RGB}{49, 27, 146}
\definecolor{deeppurpleAI}{RGB}{179, 136, 255}
\definecolor{deeppurpleAII}{RGB}{124, 77, 255}
\definecolor{deeppurpleAIV}{RGB}{101, 31, 255}
\definecolor{deeppurpleAVII}{RGB}{98, 0, 234}

\definecolor{indigoOV}{RGB}{232, 234, 246}
\definecolor{indigoI}{RGB}{197, 202, 233}
\definecolor{indigoII}{RGB}{159, 168, 218}
\definecolor{indigoIII}{RGB}{121, 134, 203}
\definecolor{indigoIV}{RGB}{92, 107, 192}
\definecolor{indigoV}{RGB}{63, 81, 181}
\definecolor{indigoVI}{RGB}{57, 73, 171}
\definecolor{indigoVII}{RGB}{48, 63, 159}
\definecolor{indigoVIII}{RGB}{40, 53, 147}
\definecolor{indigoIX}{RGB}{26, 35, 126}
\definecolor{indigoAI}{RGB}{140, 158, 255}
\definecolor{indigoAII}{RGB}{83, 109, 254}
\definecolor{indigoAIV}{RGB}{61, 90, 254}
\definecolor{indigoAVII}{RGB}{48, 79, 254}

\definecolor{blueOV}{RGB}{227, 242, 253}
\definecolor{blueI}{RGB}{187, 222, 251}
\definecolor{blueII}{RGB}{144, 202, 249}
\definecolor{blueIII}{RGB}{100, 181, 246}
\definecolor{blueIV}{RGB}{66, 165, 245}
\definecolor{blueV}{RGB}{33, 150, 243}
\definecolor{blueVI}{RGB}{30, 136, 229}
\definecolor{blueVII}{RGB}{25, 118, 210}
\definecolor{blueVIII}{RGB}{21, 101, 192}
\definecolor{blueIX}{RGB}{13, 71, 161}
\definecolor{blueAI}{RGB}{130, 177, 255}
\definecolor{blueAII}{RGB}{68, 138, 255}
\definecolor{blueAIV}{RGB}{41, 121, 255}
\definecolor{blueAVII}{RGB}{41, 98, 255}

\definecolor{lightblueOV}{RGB}{225, 245, 254}
\definecolor{lightblueI}{RGB}{179, 229, 252}
\definecolor{lightblueII}{RGB}{129, 212, 250}
\definecolor{lightblueIII}{RGB}{79, 195, 247}
\definecolor{lightblueIV}{RGB}{41, 182, 246}
\definecolor{lightblueV}{RGB}{3, 169, 244}
\definecolor{lightblueVI}{RGB}{3, 155, 229}
\definecolor{lightblueVII}{RGB}{2, 136, 209}
\definecolor{lightblueVIII}{RGB}{2, 119, 189}
\definecolor{lightblueIX}{RGB}{1, 87, 155}
\definecolor{lightblueAI}{RGB}{128, 216, 255}
\definecolor{lightblueAII}{RGB}{64, 196, 255}
\definecolor{lightblueAIV}{RGB}{0, 176, 255}
\definecolor{lightblueAVII}{RGB}{0, 145, 234}

\definecolor{cyanOV}{RGB}{224, 247, 250}
\definecolor{cyanI}{RGB}{178, 235, 242}
\definecolor{cyanII}{RGB}{128, 222, 234}
\definecolor{cyanIII}{RGB}{77, 208, 225}
\definecolor{cyanIV}{RGB}{38, 198, 218}
\definecolor{cyanV}{RGB}{0, 188, 212}
\definecolor{cyanVI}{RGB}{0, 172, 193}
\definecolor{cyanVII}{RGB}{0, 151, 167}
\definecolor{cyanVIII}{RGB}{0, 131, 143}
\definecolor{cyanIX}{RGB}{0, 96, 100}
\definecolor{cyanAI}{RGB}{132, 255, 255}
\definecolor{cyanAII}{RGB}{24, 255, 255}
\definecolor{cyanAIV}{RGB}{0, 229, 255}
\definecolor{cyanAVII}{RGB}{0, 184, 212}

\definecolor{tealOV}{RGB}{224, 242, 241}
\definecolor{tealI}{RGB}{178, 223, 219}
\definecolor{tealII}{RGB}{128, 203, 196}
\definecolor{tealIII}{RGB}{77, 182, 172}
\definecolor{tealIV}{RGB}{38, 166, 154}
\definecolor{tealV}{RGB}{0, 150, 136}
\definecolor{tealVI}{RGB}{0, 137, 123}
\definecolor{tealVII}{RGB}{0, 121, 107}
\definecolor{tealVIII}{RGB}{0, 105, 92}
\definecolor{tealIX}{RGB}{0, 77, 64}
\definecolor{tealAI}{RGB}{167, 255, 235}
\definecolor{tealAII}{RGB}{100, 255, 218}
\definecolor{tealAIV}{RGB}{29, 233, 182}
\definecolor{tealAVII}{RGB}{0, 191, 165}

\definecolor{greenOV}{RGB}{232, 245, 233}
\definecolor{greenI}{RGB}{200, 230, 201}
\definecolor{greenII}{RGB}{165, 214, 167}
\definecolor{greenIII}{RGB}{129, 199, 132}
\definecolor{greenIV}{RGB}{102, 187, 106}
\definecolor{greenV}{RGB}{76, 175, 80}
\definecolor{greenVI}{RGB}{67, 160, 71}
\definecolor{greenVII}{RGB}{56, 142, 60}
\definecolor{greenVIII}{RGB}{46, 125, 50}
\definecolor{greenIX}{RGB}{27, 94, 32}
\definecolor{greenAI}{RGB}{185, 246, 202}
\definecolor{greenAII}{RGB}{105, 240, 174}
\definecolor{greenAIV}{RGB}{0, 230, 118}
\definecolor{greenAVII}{RGB}{0, 200, 83}

\definecolor{lightgreenOV}{RGB}{241, 248, 233}
\definecolor{lightgreenI}{RGB}{220, 237, 200}
\definecolor{lightgreenII}{RGB}{197, 225, 165}
\definecolor{lightgreenIII}{RGB}{174, 213, 129}
\definecolor{lightgreenIV}{RGB}{156, 204, 101}
\definecolor{lightgreenV}{RGB}{139, 195, 74}
\definecolor{lightgreenVI}{RGB}{124, 179, 66}
\definecolor{lightgreenVII}{RGB}{104, 159, 56}
\definecolor{lightgreenVIII}{RGB}{85, 139, 47}
\definecolor{lightgreenIX}{RGB}{51, 105, 30}
\definecolor{lightgreenAI}{RGB}{204, 255, 144}
\definecolor{lightgreenAII}{RGB}{178, 255, 89}
\definecolor{lightgreenAIV}{RGB}{118, 255, 3}
\definecolor{lightgreenAVII}{RGB}{100, 221, 23}

\definecolor{limeOV}{RGB}{249, 251, 231}
\definecolor{limeI}{RGB}{240, 244, 195}
\definecolor{limeII}{RGB}{230, 238, 156}
\definecolor{limeIII}{RGB}{220, 231, 117}
\definecolor{limeIV}{RGB}{212, 225, 87}
\definecolor{limeV}{RGB}{205, 220, 57}
\definecolor{limeVI}{RGB}{192, 202, 51}
\definecolor{limeVII}{RGB}{175, 180, 43}
\definecolor{limeVIII}{RGB}{158, 157, 36}
\definecolor{limeIX}{RGB}{130, 119, 23}
\definecolor{limeAI}{RGB}{244, 255, 129}
\definecolor{limeAII}{RGB}{238, 255, 65}
\definecolor{limeAIV}{RGB}{198, 255, 0}
\definecolor{limeAVII}{RGB}{174, 234, 0}

\definecolor{yellowOV}{RGB}{255, 253, 231}
\definecolor{yellowI}{RGB}{255, 249, 196}
\definecolor{yellowII}{RGB}{255, 245, 157}
\definecolor{yellowIII}{RGB}{255, 241, 118}
\definecolor{yellowIV}{RGB}{255, 238, 88}
\definecolor{yellowV}{RGB}{255, 235, 59}
\definecolor{yellowVI}{RGB}{253, 216, 53}
\definecolor{yellowVII}{RGB}{251, 192, 45}
\definecolor{yellowVIII}{RGB}{249, 168, 37}
\definecolor{yellowIX}{RGB}{245, 127, 23}
\definecolor{yellowAI}{RGB}{255, 255, 141}
\definecolor{yellowAII}{RGB}{255, 255, 0}
\definecolor{yellowAIV}{RGB}{255, 234, 0}
\definecolor{yellowAVII}{RGB}{255, 214, 0}

\definecolor{amberOV}{RGB}{255, 248, 225}
\definecolor{amberI}{RGB}{255, 236, 179}
\definecolor{amberII}{RGB}{255, 224, 130}
\definecolor{amberIII}{RGB}{255, 213, 79}
\definecolor{amberIV}{RGB}{255, 202, 40}
\definecolor{amberV}{RGB}{255, 193, 7}
\definecolor{amberVI}{RGB}{255, 179, 0}
\definecolor{amberVII}{RGB}{255, 160, 0}
\definecolor{amberVIII}{RGB}{255, 143, 0}
\definecolor{amberIX}{RGB}{255, 111, 0}
\definecolor{amberAI}{RGB}{255, 229, 127}
\definecolor{amberAII}{RGB}{255, 215, 64}
\definecolor{amberAIV}{RGB}{255, 196, 0}
\definecolor{amberAVII}{RGB}{255, 171, 0}

\definecolor{orangeOV}{RGB}{255, 243, 224}
\definecolor{orangeI}{RGB}{255, 224, 178}
\definecolor{orangeII}{RGB}{255, 204, 128}
\definecolor{orangeIII}{RGB}{255, 183, 77}
\definecolor{orangeIV}{RGB}{255, 167, 38}
\definecolor{orangeV}{RGB}{255, 152, 0}
\definecolor{orangeVI}{RGB}{251, 140, 0}
\definecolor{orangeVII}{RGB}{245, 124, 0}
\definecolor{orangeVIII}{RGB}{239, 108, 0}
\definecolor{orangeIX}{RGB}{230, 81, 0}
\definecolor{orangeAI}{RGB}{255, 209, 128}
\definecolor{orangeAII}{RGB}{255, 171, 64}
\definecolor{orangeAIV}{RGB}{255, 145, 0}
\definecolor{orangeAVII}{RGB}{255, 109, 0}

\definecolor{deeporangeOV}{RGB}{251, 233, 231}
\definecolor{deeporangeI}{RGB}{255, 204, 188}
\definecolor{deeporangeII}{RGB}{255, 171, 145}
\definecolor{deeporangeIII}{RGB}{255, 138, 101}
\definecolor{deeporangeIV}{RGB}{255, 112, 67}
\definecolor{deeporangeV}{RGB}{255, 87, 34}
\definecolor{deeporangeVI}{RGB}{244, 81, 30}
\definecolor{deeporangeVII}{RGB}{230, 74, 25}
\definecolor{deeporangeVIII}{RGB}{216, 67, 21}
\definecolor{deeporangeIX}{RGB}{191, 54, 12}
\definecolor{deeporangeAI}{RGB}{255, 158, 128}
\definecolor{deeporangeAII}{RGB}{255, 110, 64}
\definecolor{deeporangeAIV}{RGB}{255, 61, 0}
\definecolor{deeporangeAVII}{RGB}{221, 44, 0}

\definecolor{brownOV}{RGB}{239, 235, 233}
\definecolor{brownI}{RGB}{215, 204, 200}
\definecolor{brownII}{RGB}{188, 170, 164}
\definecolor{brownIII}{RGB}{161, 136, 127}
\definecolor{brownIV}{RGB}{141, 110, 99}
\definecolor{brownV}{RGB}{121, 85, 72}
\definecolor{brownVI}{RGB}{109, 76, 65}
\definecolor{brownVII}{RGB}{93, 64, 55}
\definecolor{brownVIII}{RGB}{78, 52, 46}
\definecolor{brownIX}{RGB}{62, 39, 35}

\definecolor{grayOV}{RGB}{250, 250, 250}
\definecolor{grayI}{RGB}{245, 245, 245}
\definecolor{grayII}{RGB}{238, 238, 238}
\definecolor{grayIII}{RGB}{224, 224, 224}
\definecolor{grayIV}{RGB}{189, 189, 189}
\definecolor{grayV}{RGB}{158, 158, 158}
\definecolor{grayVI}{RGB}{117, 117, 117}
\definecolor{grayVII}{RGB}{97, 97, 97}
\definecolor{grayVIII}{RGB}{66, 66, 66}
\definecolor{grayIX}{RGB}{33, 33, 33}

\definecolor{bluegrayOV}{RGB}{236, 239, 241}
\definecolor{bluegrayI}{RGB}{207, 216, 220}
\definecolor{bluegrayII}{RGB}{176, 190, 197}
\definecolor{bluegrayIII}{RGB}{144, 164, 174}
\definecolor{bluegrayIV}{RGB}{120, 144, 156}
\definecolor{bluegrayV}{RGB}{96, 125, 139}
\definecolor{bluegrayVI}{RGB}{84, 110, 122}
\definecolor{bluegrayVII}{RGB}{69, 90, 100}
\definecolor{bluegrayVIII}{RGB}{55, 71, 79}
\definecolor{bluegrayIX}{RGB}{38, 50, 56}
\definecolor{bluegrayX}{RGB}{17, 23, 26}

\definecolor{myACMBlue}{cmyk}{1,0.1,0,0.1}
\definecolor{myACMYellow}{cmyk}{0,0.16,1,0}
\definecolor{myACMOrange}{cmyk}{0,0.42,1,0.01}
\definecolor{myACMRed}{cmyk}{0,0.90,0.86,0}
\definecolor{myACMLightBlue}{cmyk}{0.49,0.01,0,0}
\definecolor{myACMGreen}{cmyk}{0.20,0,1,0.19}
\definecolor{myACMPurple}{cmyk}{0.55,1,0,0.15}
\definecolor{myACMDarkBlue}{cmyk}{1,0.58,0,0.21}

\newcommand{\link}[1]{{\href{#1}{\color{blueVI}\textbf{\texttt{#1}}}}}

\crefname{figure}{fig.}{fig.}
\Crefname{figure}{Fig.}{Fig.}
\crefname{equation}{eq.}{eq.}
\Crefname{equation}{Eq.}{Eq.}
\crefname{section}{\S}{\S}
\creflabelformat{equation}{#2\textup{#1}#3}

\newcommand{\figpart}[1]{\textcolor{myACMPurple}{#1}}
\hypersetup{colorlinks,
      linkcolor=myACMPurple,
      citecolor=myACMPurple,
      urlcolor=black,
      filecolor=myACMDarkBlue}

\newcommand{\tool}{\textsc{TimberTrek}}

\newcommand{\summaryview}{\textit{Rashomon Overview}}
\newcommand{\treeview}{\textit{Tree Window}}
\newcommand{\treeviews}{\textit{Tree Windows}}
\newcommand{\searchview}{\textit{Search Panel}}
\newcommand{\favoriteview}{\textit{Favorite Panel}}

\definecolor{soulorange}{RGB}{255, 212, 153}
\definecolor{soulgray}{RGB}{220, 220, 220}
\definecolor{soulgraylight}{RGB}{235, 235, 235}
\definecolor{soulred}{RGB}{252, 217, 218}
\definecolor{soulbluelight}{RGB}{208, 233, 253}
\definecolor{souldorangelight}{RGB}{254, 234, 212}
\colorlet{soulblue}{blueV!30}

\newcommand{\inlinefig}[2]{\protect\includegraphics[align=c, height=#1pt]{figures/#2}}

\definecolor{tagbordercolor}{rgb}{0.8, 0.8, 0.8}
\definecolor{tagbgcolor}{rgb}{0.9, 0.9, 0.9}

\newtcbox{\tagg}{nobeforeafter, colframe=tagbordercolor,
colback=tagbgcolor, boxrule=0.5pt, arc=1pt,
  boxsep=0pt,left=2pt,right=2pt,top=1.5pt,bottom=2pt,tcbox raise base}

\definecolor{lightgray}{RGB}{247, 247, 247}
\definecolor{midgray}{RGB}{179, 179, 179}

\newtcbox{\featuretag}{on line,
  colframe=midgray,colback=lightgray,
  boxrule=0.5pt,arc=2pt,boxsep=0pt,left=2pt,right=1pt,top=1pt,bottom=1pt
}
\definecolor{tagbgcolor}{rgb}{1, 1, 1}

\definecolor{boxyellow}{RGB}{206, 171, 1}
\definecolor{boxgreen}{RGB}{14, 152, 136}
\definecolor{boxblue}{RGB}{77, 167, 223}

\setul{0.45ex}{0.2ex}

\begin{document}

\firstsection{Introduction}

\maketitle

It is essential to understand how machine learning (ML) models make predictions in high-stakes settings such as healthcare, finance, and criminal justice.
Researchers have made great strides in developing interpretable models~\cite[e.g.,][]{linGeneralizedScalableOptimal2020,caruanaIntelligibleModelsHealthCare2015,ustunLearningOptimizedRisk2019} that perform competitively with state-of-the-art black-box models yet have transparent and simple structures~\cite{wangPursuitInterpretableFair2022,changHowInterpretableTrustworthy2021}.
Some recent research focuses on \textit{operationalizing} interpretability---leveraging an understanding of the domain to create more responsible and trustworthy ML systems~\cite{wangInterpretabilityThenWhat2022a, noriAccuracyInterpretabilityDifferential2021}.\looseness=-1

To help ML practitioners build trustworthy models, researchers have recently developed a technique to generate the full set of almost-optimal sparse decision trees \cite{xinExploringWholeRashomon2022}.
This set of high-performing models is called the \textit{Rashomon set}~\cite{fisherAllModelsAre2019, costonCharacterizingFairnessSet2021}, named after the \textit{Rashomon effect} in statistics~\cite{breimanStatisticalModelingTwo2001}.
A Rashomon set of sparse decision trees can have thousands of inherently interpretable and almost-equally accurate models~\cite{xinExploringWholeRashomon2022}, providing opportunities for users to choose ones that best align with their knowledge and needs (e.g., fairness, monotonicity, simplicity)~\cite{dongExploringCloudVariable2020,costonCharacterizingFairnessSet2021}.
However, the large size of the Rashomon set and diversity of models within it pose challenges for users wishing to effectively explore the set and compare these accurate models~\cite {rudinInterpretableMachineLearning2022}.
To tackle this critical challenge, we
\textbf{contribute}:\looseness=-1

\begin{figure}[tb]
  \setlength{\belowcaptionskip}{-5pt}
  \setlength{\abovecaptionskip}{5pt}
  \includegraphics[width=\columnwidth]{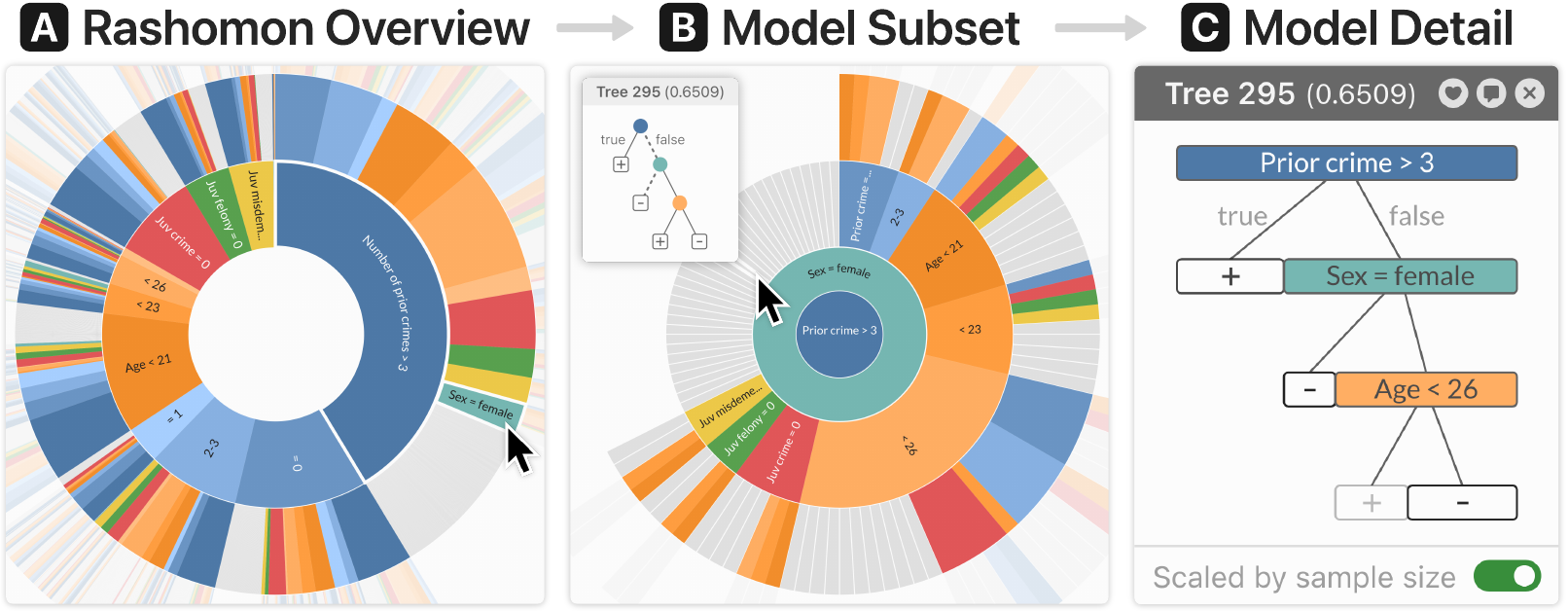}
  \caption{
    \tool{}'s tightly integrated views enable users to characterize and curate model candidates by seamlessly traversing across abstraction levels.
    \textbf{(A)} The \summaryview{} summarizes the entire Rashomon set;
    \textbf{(B)} zoomable Sunburst enables users to focus on a subset of models with similar prediction patterns; and
    \textbf{(C)} the \treeview{} presents the details of a selected decision tree.
  }
  \label{fig:transition}
  \setlength{\belowcaptionskip}{0pt}
  \setlength{\abovecaptionskip}{10pt}
\end{figure}

\begin{itemize}[topsep=1pt, itemsep=0mm, parsep=3pt, leftmargin=9pt]

\item \textbf{\tool{}, the first interactive visualization tool} that empowers domain experts and data scientists to easily explore the Rashomon set of sparse decision trees and curate models with desired properties.
\autoref{fig:teaser} shows an example of \tool{} in action, helping users explore 5384 decision paths from 1365 trees for recidivism risk assessment~\cite{larsonHowWeAnalyzed2016}.
Advancing over prior visual analytics tools designed for interpretable ML models~\cite[e.g.,][]{wangInterpretabilityThenWhat2022a,hohmanGamutDesignProbe2019}, our tool overcomes unique design challenges identified from a literature review of recent work in ML interpretability~(\autoref{sec:goals}).

\item \textbf{Novel interactive system design} that leverages Sunburst~\cite{staskoFocusContextDisplay2000} to summarize the entire Rashomon set at scale by organizing decision trees based on their decision paths~(\autoref{fig:teaser}\figpart{A}).
Through animation and \textit{focus+context}~\cite{cardReadingsInformationVisualization1999} techniques~(\autoref{sec:interface}), our tool enables users to seamlessly traverse the full spectrum of abstraction levels: from the highest-level Sunburst overview~(\autoref{fig:transition}\figpart{A}), to intermediate levels of model subsets with similar prediction patterns~(\autoref{fig:transition}\figpart{B}), to the lowest-level node-link representation of individual trees~(\autoref{fig:transition}\figpart{C}).
Two usage scenarios highlight how our tool can help users curate models with desired properties~(\autoref{sec:scenario}).

\item \textbf{An open-source\footnote{\tool{} code: \link{https://github.com/poloclub/timbertrek}} and web-based implementation} that broadens people's access to trustworthy ML techniques~(\autoref{sec:interface:implementation}).
We develop \tool{} with modern web technologies so that anyone can access our tool directly in their web browsers and computational notebooks.
For a demo video of \tool{}, visit \link{https://youtu.be/3eGqTmsStJM}.
\end{itemize}

\noindent We hope our work helps democratize cutting-edge responsible ML techniques as well as inspires and informs future work in human-AI interaction and visual analytics for interpretable ML. %
\section{Background \& Related Work}

\textbf{Decision trees and Rashomon set.}
Decision trees have been popular for more than half a century due to their accuracy and flexibility~\cite{lohFiftyYearsClassification2014, morganProblemsAnalysisSurvey1963}.
These predictive models have a tree structure, where each branch node assesses a condition, and each leaf makes a prediction.
With modern optimization techniques~\cite{linGeneralizedScalableOptimal2020, huOptimalSparseDecision2019}, sparse decision trees---decision trees using a small set of features---have been gaining popularity in health care and criminal justice~\cite{rudinStopExplainingBlack2019}, as these models are not only accurate but also simple enough to be memorized by humans~\cite{rudinInterpretableMachineLearning2022}.
To help users explore diverse and accurate models and eventually find ones that they can trust, recent researchers have developed an algorithm to generate the whole Rashomon set of sparse decision trees~\cite{xinExploringWholeRashomon2022}.
Given a dataset with binary features and binary labels, two hyperparameters (sparsity penalty $\lambda$ and loss tolerance $\epsilon$), this algorithm will find all binary decision trees with at most $\epsilon \times \ell$ loss on the training data, where $\ell$ is the loss of the optimal tree.
To characterize a Rashomon set, researchers have proposed visualization techniques to study the model construction process~\cite{kisselForwardStabilityModel2021} and feature importance~\cite{dongExploringCloudVariable2020}.
Different from these techniques, \tool{} is the first interactive tool that summarizes a Rashomon set with varying levels of abstraction and enables users to curate good models.\looseness=-1

\textbf{Visual analytics for model selection.}
Iterating and selecting good models is a critical part of ML workflows~\cite{amershiSoftwareEngineeringMachine2019,amershiPowerPeopleRole2014}.
Visual analytics tools have shown great success in facilitating model selection~\cite[e.g.,][]{bradelMultimodelSemanticInteraction2014,cashmanAblateVariateContemplate2020, chatzimparmpasStackGenVisAlignmentData2021}, as they enable the integration of domain knowledge in model development~\cite{streebTaskbasedVisualInteractive2021, chatzimparmpasStateArtEnhancing2020}.
For example, BEAMES~\cite{dasBEAMESInteractiveMultimodel2019} is an interactive tool that helps domain experts impose model constraints and search for linear models that meet specified constraints.
Researchers have also developed visual analytics tools for interpreting tree-based models~\cite[e.g.,][]{chatzimparmpasVisRulerVisualAnalytics2022, zhaoIForestInterpretingRandom2019, liVisualAnalyticsSystem2020, netoExplainableMatrixVisualization2021}.
However, these tools focus on understanding single tree ensembles (i.e., random forest~\cite{breimanRandomForests2001} and gradient-boosted trees~\cite{friedmanGreedyFunctionApproximation2001}) instead of choosing from a collection of standalone decision trees.
Possibly closest in spirit to \tool{} are \textsc{TreePOD}~\cite{muhlbacherTreePODSensitivityAwareSelection2018} and a system designed by Padua et al.~\cite{paduaInteractiveExplorationParameter2014}.
Both tools guide users to select satisfactory decision trees by tuning the parameters of decision tree algorithms.
In contrast, our tool visualizes the complete Rashomon set of sparse decision trees with different levels of abstraction. For \tool{}, every tree produced has both performance and interpretability guarantees.

\section{Design Goals}
\label{sec:goals}

Through synthesizing recent work in interpretable ML and ML workflows, we identify four design goals~(\ref{item:g1}--\ref{item:g4}) for \tool{}.

\begin{enumerate}[topsep=2pt, itemsep=0mm, parsep=3pt, leftmargin=19pt, label=\textbf{G\arabic*.}, ref=G\arabic*]
  \item \label{item:g1}
  \textbf{Visual Summary of the whole Rashomon set.}
  Depending on hyperparameters $\lambda$ and $\epsilon$, the Rashomon set's size can vary from hundreds to tens of thousands~\cite{xinExploringWholeRashomon2022}, posing challenges for users to explore models in this set~\cite{rudinInterpretableMachineLearning2022}.
  Therefore, we aim to design scalable visualizations to summarize a large number of sparse decision trees to help users gain a better understanding of the landscape of the Rashomon set~(\autoref{sec:interface:summarize}).

  \item \label{item:g2}
  \textbf{Fluid transition between different levels of abstraction.}
  To curate models, users need to characterize all model candidates~\cite{rudinInterpretableMachineLearning2022}, identify important features~\cite{fisherAllModelsAre2019,dongExploringCloudVariable2020}, and compare individual models~\cite{amershiPowerPeopleRole2014}.
  Therefore, we would like to design a focus + context display to help users easily connect the Rashomon set landscape to individual models~(\autoref{sec:interface:summarize}).
  In addition, we would like to design query mechanisms to enable users to quickly pinpoint models with desirable properties~(\autoref{sec:interface:search}, \autoref{sec:interface:curate}).

  \item \label{item:g3}
  \textbf{Model comparison.}
  Model selection often requires model comparison~\cite{amershiPowerPeopleRole2014,costonCharacterizingFairnessSet2021}.
  In our case, all models are interpretable and have similar accuracies; thus, we would like \tool{} to help users compare decision tree structures and prediction patterns~(\autoref{sec:interface:compare}).
  As each model in the Rashomon set provides a \textit{different} and \textit{incomplete} explanation of the real-word phenomena~\cite{breimanStatisticalModelingTwo2001}, comparison can also help users gain insights into patterns within the full set of reasonable possibilities.\looseness=-1

  \item \label{item:g4}
  \textbf{Fit into model development workflows.}
  Computational notebooks, such as Jupyter Notebook~\cite{kluyverJupyterNotebooksPublishing2016}, have revolutionalized how ML practitioners develop models~\cite{perkelReactiveReproducibleCollaborative2021}.
  To make model curation accessible and fit into the current workflows, we would like \tool{} to work in both web browsers and computational notebooks.
  Finally, we open-source our implementation to support future design, research, and development of visual analytics tools for interpretable ML~(\autoref{sec:interface:implementation}).

\end{enumerate} %
\section{Visualization Interface of \tool{}}
\label{sec:interface}

\definecolor{tempblue}{HTML}{4E79A7}
\definecolor{temporange}{HTML}{F28E2C}
\definecolor{temporangeii}{HTML}{FDAF6E}
\definecolor{temporangeiv}{HTML}{FFD8B8}
\definecolor{tempblueii}{HTML}{6B95C4}
\definecolor{tempblueiv}{HTML}{88B1E2}
\definecolor{tempteal}{HTML}{76B7B2}
\definecolor{tempred}{HTML}{E15759}

Following the design goals, \tool{}~(\autoref{fig:teaser}) tightly integrates four components: the \summaryview{} providing a hierarchical overview of all decision trees in a Rashomon set~(\autoref{sec:interface:summarize}), the \searchview{} enabling users to find trees with desirable properties~(\autoref{sec:interface:search}), \treeviews{} showing details of selected decision trees~(\autoref{sec:interface:compare}), and the \favoriteview{} documenting curated trees~(\autoref{sec:interface:curate}).\looseness=-1

\subsection{Summarizing the Whole Rashomon Set}
\label{sec:interface:summarize}

A Rashomon set can contain thousands of sparse binary decision trees.
Consider criminal recidivism prediction as an example~\cite{larsonHowWeAnalyzed2016}; each branch node in a decision tree assesses a feature condition~(e.g., \textcolor{tempred}{\textbf{juvenile crime = 0}}), and a leaf node makes a prediction (e.g., the subject is likely to reoffend in two years).
Decision rules~\cite{changHowInterpretableTrustworthy2021,streebTaskbasedVisualInteractive2021} can be represented as a path from the root to a leaf.
For example, the left-most path of \texttt{Tree} \texttt{1071}~(\autoref{fig:favorite}\figpart{A}) represents ``IF \textcolor{tempred}{\textbf{(juvenile crime = 0)}} AND \textcolor{tempblue}{\textbf{(prior crime \texttt{>} 3)}} THEN (the subject is likely to re-offend).''
To help users characterize the Rashomon set, we can organize decision trees based on their decision rules.

\textbf{Sunburst overview.} We first construct a trie of decision rules by extracting decision paths from all leaves of all trees in the Rashomon set~(\ref{item:g1}).
A trie leaf is a decision tree that contains a decision rule matching the leaf's ancestors.
We use Sunburst~\cite{staskoFocusContextDisplay2000} to visualize the trie in the \summaryview{}~(\autoref{fig:teaser}\figpart{{A}}).
The Sunburst consists of $h$ concentric rings, where $h$ is the height of the decision rule trie (i.e., the maximal tree height in the Rashomon set).
A ring corresponds to a level in the trie, and it is segmented into annular sectors, where each sector represents a split condition used in decision rules.
Each sector cumulatively starts a subtrie, where the children are conditions used in the next level of the trie.
A sector's size is proportional to the number of its descendant decision trees.

\textbf{Color encoding.}
We use the Hue-chroma-luminance (HCL)~\cite{zeileisEscapingRGBlandSelecting2009} color space to represent the sector's color.
Decision tree methods often binarize continuous and categorical features into multiple split conditions covering various ranges~\cite{huOptimalSparseDecision2019,linGeneralizedScalableOptimal2020}.
Therefore, we use different hue values to represent different features (e.g., \textcolor{tempblue}{\textbf{prior crime}}, \textcolor{temporange}{\textbf{age}}, \textcolor{tempteal}{\textbf{sex}}), and luminance values to represent different ranges of the same feature (e.g., \textcolor{tempblue}{\textbf{prior \texttt{>} 3}}, \textcolor{tempblueii}{\textbf{prior \texttt{=} 0}}, \textcolor{tempblueiv}{\textbf{prior \texttt{=} 1}}).
We use \textcolor{grayV}{\textbf{gray}} to encode \textcolor{grayV}{\textbf{leaf sectors}}: a leaf of the trie indicates the end of a decision rule, and each leaf links to a decision tree that contains this decision rule.
Finally, we group sectors using the same feature and sort sectors based on the number of trees within the Rashomon set.

\setlength{\columnsep}{5pt}%
 \setlength{\intextsep}{0pt}%
 \begin{wrapfigure}{R}{0.17\textwidth}
   \vspace{0pt}
   \centering
   \includegraphics[width=0.17\textwidth]{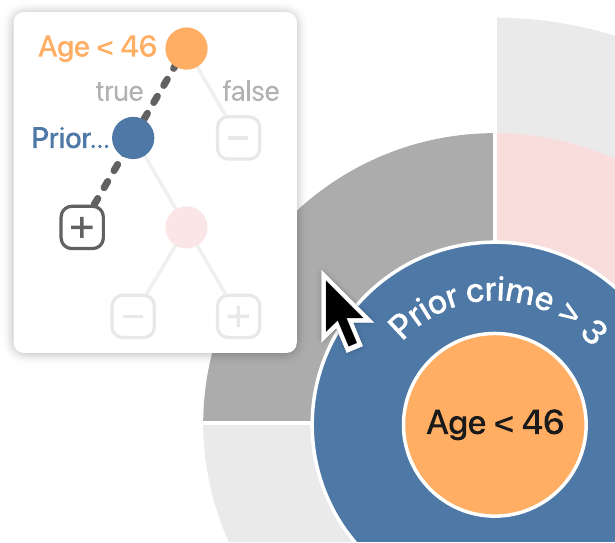}
   \vspace{-10pt}
 \end{wrapfigure}
\textbf{Model exploration.}
The \summaryview{} not only summarizes all decision trees in the Rashomon set~(\ref{item:g1}) but also highlights \textit{feature importance} through sectors' size, as important features are often used by many accurate models thus yield larger sectors~\cite{dongExploringCloudVariable2020}.
To help users further explore decision trees with similar prediction patterns, the \summaryview{} provides smooth transitions between different levels of Rashomon set abstraction~(\ref{item:g2}).
When users click a sector, the Sunburst's root switches to the selected split condition and only displays its descendants~(\autoref{fig:transition}\figpart{B})---helping users focus on an interesting subtrie.
When users hover over a \textcolor{grayV}{\textbf{leaf sector}}, our tool shows the corresponding decision tree as a link-node diagram~\cite{nguyenVisualizationToolInteractive2000}~(shown on the right), where the selected decision rule is highlighted as animated dash lines and binary outputs as \inlinefig{8}{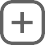} and \inlinefig{8}{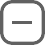}.
Users can control how many levels to display in Sunburst via the depth panel, whose colors match clicked sectors~(\autoref{fig:teaser}\figpart{A-top left}).\looseness=-1

\subsection{Searching Models with Desirable Properties}
\label{sec:interface:search}

The \summaryview{} provides a bird's-eye view, enabling users to follow different decision paths to explore different subsets of decision trees in the Rashomon Set.
To allow users to quickly pinpoint trees with desirable properties~(\ref{item:g2}), the \searchview{}~(\autoref{fig:teaser}\figpart{C}) offers a suite of filtering panels to control what trees to display in the \summaryview{}.
For example, with the accuracy and minimum sample leaf size sliders, users can focus on trees with desired accuracy and robustness.
Similarly, users can use checkboxes to filter models by tree height and the use of specific features~(\autoref{fig:compas}\figpart{A}).

\subsection{Comparing Individual Decision Trees}
\label{sec:interface:compare}

\begin{figure}[tb]
  \setlength{\belowcaptionskip}{-5pt}
  \setlength{\abovecaptionskip}{5pt}
  \includegraphics[width=\columnwidth]{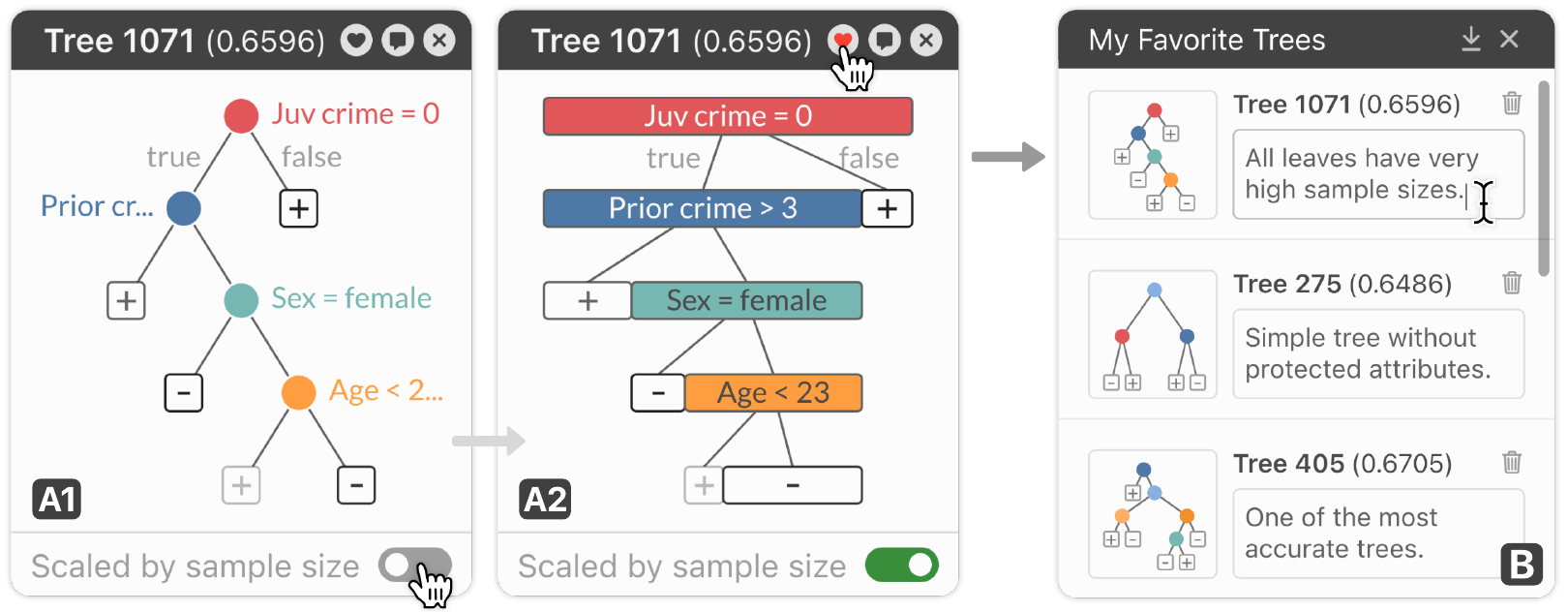}
  \caption{
    \tool{} helps users inspect and curate decision trees.
    \textbf{(A1)} The \treeview{} visualizes a decision tree's prediction pattern via a node-link diagram;
    \textbf{(A2)} a funnel-like complementary view scales tree nodes by their sample sizes.
    \textbf{(B)} The \favoriteview{} allows users to keep track of bookmarked models with curation documentation.
  }
  \label{fig:favorite}
  \setlength{\belowcaptionskip}{0pt}
  \setlength{\abovecaptionskip}{10pt}
\end{figure}

The combination of \summaryview{} and \searchview{} provides users with a searchable directory of decision trees.
After identifying interesting model candidates, users often need to compare them to identify ones suitable for practical use~(\ref{item:g3}).
When a user clicks a \textcolor{grayV}{\textbf{leaf sector}}, a \treeview{} appears, visualizing the corresponding decision tree as a node-link diagram~(\autoref{fig:favorite}\figpart{-A1}).
We use opacity to encode a leaf node's accuracy, allowing users to quickly inspect a tree's purity and prediction confidence~\cite{breimanClassificationRegressionTrees1984}.
This window is repositionable through dragging, and there can be multiple windows open at once---a user can create \treeviews{} for all interesting model candidates and easily compare their structures and prediction patterns side by side.
When comparing decision trees, users are also interested in the sample sizes of nodes in addition to the tree structure~\cite{streebTaskbasedVisualInteractive2021}.
Therefore, when a user toggles the sample-size switch, the \treeview{} transitions the node width to represent the percentage of training samples that fall into each split condition (branch node) or prediction (leaf node)~(\autoref{fig:favorite}\figpart{-A2}).
This novel funnel-like node-link diagram can help users quickly identify important nodes and evaluate model robustness via the node sample sizes~\cite{breimanClassificationRegressionTrees1984}.

\subsection{Curating Trustworthy Models}
\label{sec:interface:curate}

The goal of \tool{} is to empower users to explore and curate decision trees for practical use.
Once a user has identified a satisfactory model, they can click the heart button~\inlinefig{8}{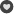} in the \treeview{} to bookmark a decision tree.
Bookmarked trees appear in the \favoriteview{}~(\autoref{fig:favorite}\figpart{B}).
To help users track their reasons and contexts for choosing a particular decision tree, the \favoriteview{} allows users to attach a comment to each bookmarked model.
Alternatively, users can also click the comment button~\inlinefig{8}{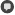} to add comments directly in the \treeview{}~(\autoref{fig:compas}\figpart{C}).
These comments allow users to continue their model curation in the future~(\ref{item:g4}), help ML auditors audit models before deployment~\cite{wangInterpretabilityThenWhat2022a,amershiGuidelinesHumanAIInteraction2019}, and help improve ML transparency regarding the model development process~\cite{hancox-liRobustnessMachineLearning2020}.
Finally, users can click the save button~\inlinefig{8}{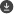} to export bookmarked decision trees with curation comments; \tool{}'s companion package provides an API to load and deploy saved models.

\subsection{Accessible, Open-source Implementation}
\label{sec:interface:implementation}
\tool{} is a web-based interactive visualization tool built with \textit{D3.js}~\cite{bostockDataDrivenDocuments2011}: users can access our tool with any web browser or directly in computational notebooks.
To promote the accessibility of our tool, and strongly align with our VIS community's open science practice, we have released \tool{} on the Python Package Index (PyPI),\footnote{PyPI repository: \link{https://pypi.org/project/timbertrek/}}  so that users can easily install our tool and integrate it into their ML development workflows~(\ref{item:g4}).
We have also open sourced our implementation: future researchers can quickly adapt our design to other forms of model curation.

\clearpage %
\section{Usage Scenarios}
\label{sec:scenario}

\begin{figure}[tb]
  \centering
  \setlength{\belowcaptionskip}{-11pt}
  \setlength{\abovecaptionskip}{7pt}
  \includegraphics[height=181.5pt]{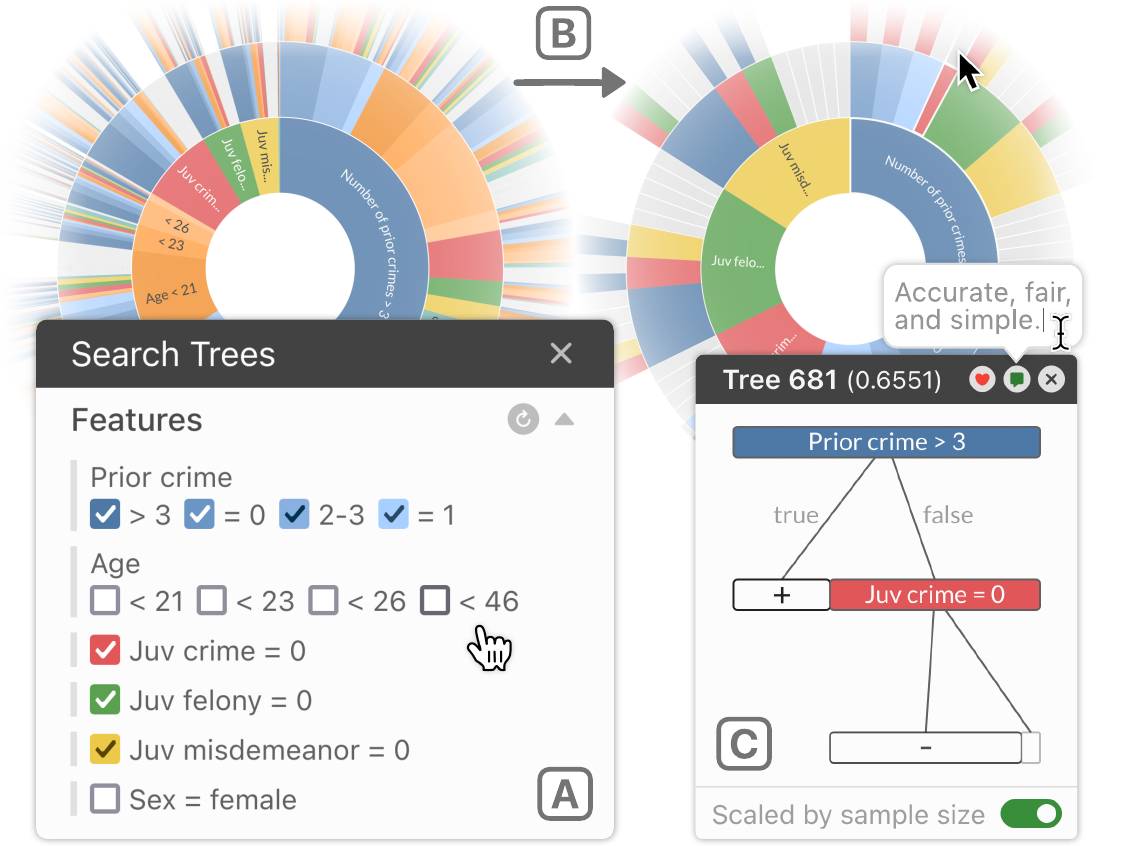}
  \caption{
    With \tool{}, users can easily search for models with desirable properties.
    Here, in the example of criminal recidivism assessment,
    \textbf{(A)} the \searchview{} enables users to query models that do not use protected attributes (e.g., age and sex);
    \textbf{(B)} the \summaryview{} animates to only display models meeting the query criteria.
    \textbf{(C)} The \treeview{} provides details of selected model query result.
  }
  \label{fig:compas}
  \setlength{\belowcaptionskip}{0pt}
  \setlength{\abovecaptionskip}{10pt}
\end{figure}

We present two hypothetical scenarios with real datasets to demonstrate how \tool{} can potentially help data scientists and domain experts gain a better understanding of the model-world relationship and curate more trustworthy models.
We generate a Rashomon set with 1,365 trees ($\lambda=0.01$, $\epsilon=1.05$) for the COMPAS recidivism dataset~\cite{larsonHowWeAnalyzed2016}~(\autoref{sec:scenario:compas}), and a Rashomon set with 911 trees ($\lambda=0.15$, $\epsilon=1.015$) for the Car Evaluation dataset~\cite{bohanecKnowledgeAcquisitionExplanation1988}~(\autoref{sec:scenario:car}).

\subsection{Discovering Fair Rules for Recidivism Assessment}
\label{sec:scenario:compas}

Mei is a data scientist who develops transparent and fair ML models to help inform judicial bail decisions.
To explore the relationship between diverse variables and the risk of criminal recidivism, she has generated a Rashomon set of sparse decision trees on past criminal recidivism data (we use COMPAS~\cite{larsonHowWeAnalyzed2016} to illustrate this scenario).
This dataset includes defendants' demographic information and criminal history; the outcome variable is binary---indicating whether a defendant is likely to reoffend in the next two years.

To understand the similarities and differences between all 1,365 about-equally accurate models from the Rashomon set, Mei loads all models into \tool{}.
The \summaryview{} visualizes all decision trees~(\autoref{fig:transition}\figpart{A}).
Inspecting the Sunburst's first ring, Mei quickly realizes \textcolor{tempblue}{\textbf{prior crime}} may be the most important feature to assess recidivism risk.
This is because the root is the most powerful node in a decision tree~\cite{breimanClassificationRegressionTrees1984}, and \textit{more than half} of models in the Rashomon set choose \textcolor{tempblue}{\textbf{prior crime}} as their roots.
Mei opens the \searchview{}~(\autoref{fig:teaser}\figpart{C}) and searches for trees that do not use \textcolor{tempblue}{\textbf{prior crime}} at any depth---there is no tree meeting this criterion.
Similarly, Mei hypothesizes \textcolor{tempteal}{\textbf{sex}} is the least important feature due to the small sizes of its sectors~(\autoref{fig:transition}\figpart{A}).
Mei's hypotheses regarding feature importance match previous study results on COMPAS~\cite{dongExploringCloudVariable2020}.\looseness=-1

Mei believes \textcolor{tempblue}{\textbf{prior crime}} is indeed an informative feature for recidivism prediction, but she notices many accurate models use sensitive features such as \textcolor{temporange}{\textbf{age}} and \textcolor{tempteal}{\textbf{sex}}.
Making bail decisions with these models could be problematic~\cite{chouldechovaFairPredictionDisparate2017}.
Therefore, Mei decides to find accurate models that do not use any sensitive features.
To do that, she uses the \searchview{} to query decision trees without any \textcolor{temporange}{\textbf{age}} or \textcolor{tempteal}{\textbf{sex}} nodes~(\autoref{fig:compas}\figpart{A}).
\tool{} finds 33 trees that meet this criterion~(\autoref{fig:compas}\figpart{B}).
After inspecting these trees by hovering over the \textcolor{grayV}{\textbf{leaf sectors}}, Mei finds her favorite model---\texttt{Tree 681} that does not use sensitive features, yields a high accuracy, and is simple enough to be memorized~(\autoref{fig:compas}\figpart{C}).
Mei adds it to her curation list and writes a short comment to document why she chooses this tree.

\begin{figure}[tb]
  \centering
  \setlength{\belowcaptionskip}{-11pt}
  \setlength{\abovecaptionskip}{7pt}
  \includegraphics[height=181.5pt]{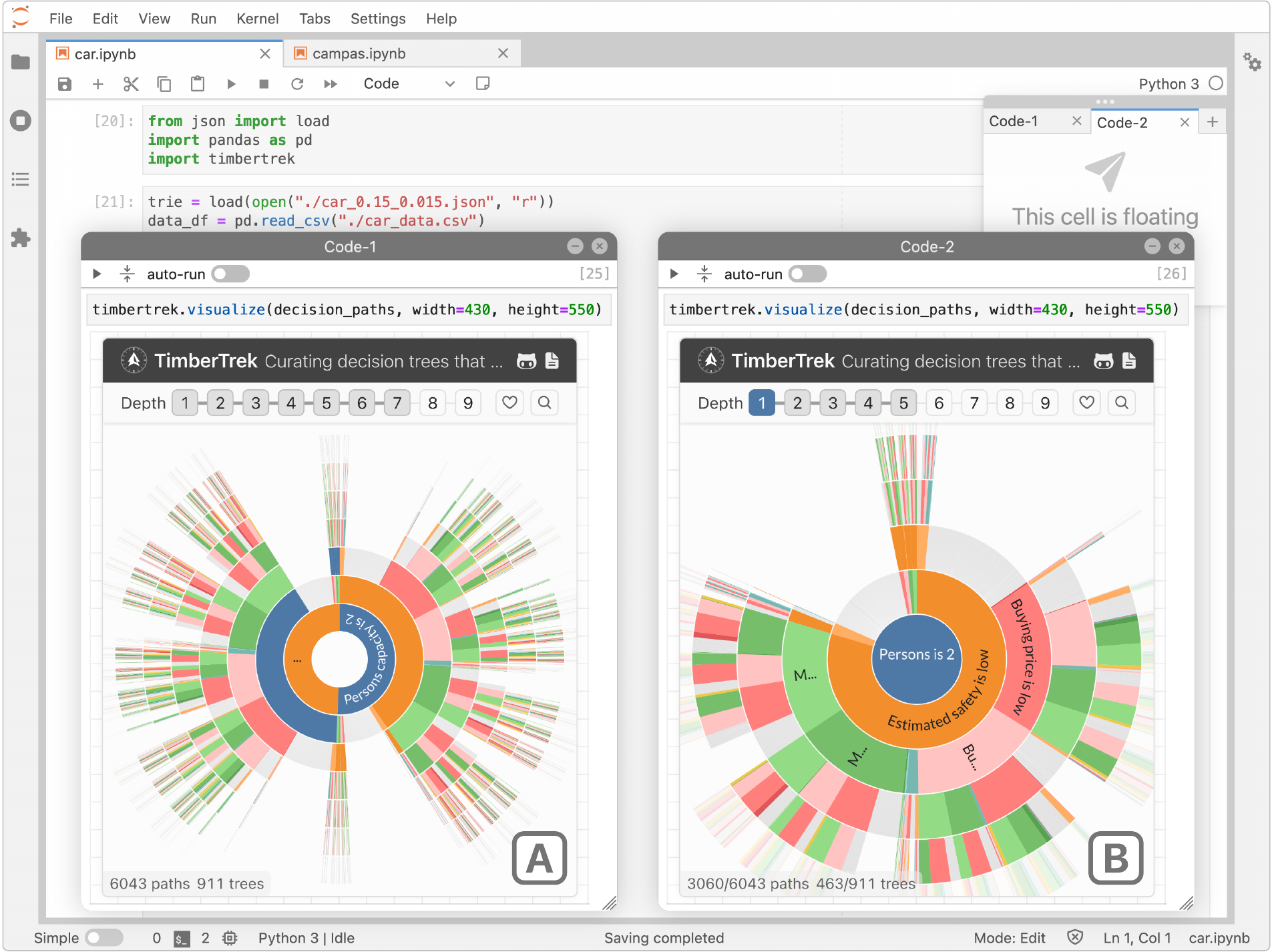}
  \caption{
    Our tool works in computational notebooks---commonly used in ML development workflows.
    With sticky cells, users can create multiple \tool{} instances and compare different Rashomon subsets side by side.
    Take Car Evaluation as an example: \textbf{(A)} reveals all trees in the Rashomon set use either \textcolor{tempblue}{\textbf{person}} or \textcolor{temporange}{\textbf{safety}} as root, and \textbf{(B)} provides an additional view to focus on trees with a \textcolor{tempblue}{\textbf{person}} root.\looseness=-1
  }
  \label{fig:notebook}
  \setlength{\belowcaptionskip}{0pt}
  \setlength{\abovecaptionskip}{10pt}
\end{figure}

\subsection{Curating Trustworthy Models for Insurance Quote}
\label{sec:scenario:car}

Robaire is an ML developer building models to help his company make automobile insurance quotes.
For accountability to end-users, his company requires all models to be transparent and easy to explain.
Therefore, Robaire decides to curate trustworthy models from the Rashomon set of sparse decision trees.
We use the Car Evaluation dataset~\cite{bohanecKnowledgeAcquisitionExplanation1988} to illustrate this scenario; the task is to predict whether a vehicle's value is acceptable with typical vehicle features.
After loading \tool{} directly in JupyterLab~(\autoref{fig:notebook}\figpart{{A}}), Robaire quickly notices all 911 accurate trees (median accuracy is \texttt{0.92})---with an even distribution---use either \textcolor{tempblue}{\textbf{person capacity}} or \textcolor{temporange}{\textbf{safety score}} as their first split.
Curious, Robaire decides to compare model structures between two subsets of trees with different roots.
He opens a second \tool{} instance in a notebook sticky cell~\cite{wangStickyLandBreakingLinear2022}, where he clicks the \textcolor{tempblue}{\textbf{blue sector}} in the first ring to focus on trees with a \textcolor{tempblue}{\textbf{person capacity}} root~(\autoref{fig:notebook}\figpart{B}).
He repeats the same process on the other cell to choose a different root.
Comparing two subsets side by side, Robaire finds almost all trees use these two features in their first two splits and use diverse combinations of other features in further levels.
Therefore, to avoid potential overfitting and keep the model simple enough for end-users to understand, Robaire eventually chooses two trees that use only \textcolor{tempblue}{\textbf{person capacity}} and \textcolor{temporange}{\textbf{safety score}} nodes.
\looseness=-1

\vspace{2pt} %
\section{Discussion \& Conclusion}

We present \tool{}, the first visualization system that summarizes the entire Rashomon set and empowers ML practitioners to explore, compare, and curate models with desired properties.
Our current prototype does not scale to multi-class classification trees, decision trees with many features (limitation of our color encoding), or trees with many levels.
However, interpretability requires decision trees to limit the number of features and levels~\cite{yuanVisualExplorationMachine2022,rudinInterpretableMachineLearning2022}.
In addition, our design principles such as focus + context, model comparison and query are generalizable to other model types.
Future researchers can use our tool as a research instrument to probe how users would select ML models when many models are approximately equally accurate.
We hope our work will inspire future research and development of tools that can empower users to interpret and trust ML technologies.\looseness=-1 %

\bibliographystyle{abbrv}

\clearpage
\acknowledgments{
  We thank anonymous reviewers for their valuable feedback. This work was supported in part by a J.P. Morgan PhD Fellowship, NSF grants IIS-1563816, CNS-1704701, NIH/NIDA grant DA054994-01, DARPA GARD, Fujitsu, gifts from Intel, NVIDIA, Bosch, Google.
}

\setlength{\bibsep}{0.29pt}
{\footnotesize
\bibliography{timbertrek}
}

\end{document}